\def\rfr#1{eq. (\ref{#1})}

\def\dert#1#2{\frac{{{d}}{#1}}{{{d}}{#2}}}              

\def\bar{\begin{eqnarray}}
\def\ear{\end{eqnarray}}

\def\eqi{\begin{equation}}
\def\eqf{\end{equation}}
\def\eqia{\begin{eqnarray}}
\def\eqfa{\end{eqnarray}}
\def\rp#1#2{{#1\over#2}}

\def\lb#1{\label{#1}}

\def\oc2{$\mathcal{O}(c^{-2})$}

\documentclass[11pt]{article}
\usepackage{amsmath,amsthm,amscd,amssymb}
\usepackage{latexsym}
\usepackage{graphicx,epsfig}

\begin{document}

\noindent{\bf \LARGE{A comment on the new non-conventional
gravitational mechanism proposed by Jaekel and Reynaud to
accommodate the Pioneer anomaly }}
\\
\\
\\
{Lorenzo Iorio}\\
{\it Viale Unit$\grave{a}$ di Italia 68, 70125\\Bari, Italy
\\tel./fax 0039 080 5443144
\\e-mail: lorenzo.iorio@libero.it}

\begin{abstract}
In this paper we put on the test the new mechanism of
gravitational origin recently put  forth by Jaekel and Reynaud in
order to explain the Pioneer anomaly in the framework of their
post-Einsteinian metric extension of general relativity. According
to such a proposal, the secular part of the anomalous acceleration
experienced by the twin spacecraft of about 1 nm s$^{-2}$ could be
caused by an extra-potential $\delta\Phi_{P}=c^2\chi r^2$, with
$\chi=4\times 10^{-8}$ AU$^{-2}$, coming from the second sector of
the considered model. When applied to the motion of the planets of
the Solar System, it would induce anomalous secular perihelion
advances which amount to  tens-hundreds of arcseconds per century
for the outer planets. As for other previously proposed
non-conventional gravitational explanations of the Pioneer
anomaly, the answer of the latest determinations of the anomalous
perihelion rates by RAS IAA is neatly and unambiguously negative.
The presence of another possible candidate to explain the Pioneer
anomaly, i.e. the extra-potential $\delta\Phi_N$, linear in
distance,  from the first sector of the Jaekel and Reynaud model,
is ruled out not only by the residuals of the optical data of the
outer planets processed with the recent RAS IAA EPM2004
ephemerides but also by those produced with other, older dynamical
theories like, e.g., the well known NASA JPL DE200.
\end{abstract}



In the framework of their post-Einsteinian metric extension of
general relativity, Jaekel and Reynaud (2006) recently put fort a
new model which amends a previous one by the same authors (Jaekel
and Reynaud 2005a; 2005b) and, among other things, yields a
possible explanation of gravitational origin of the secular part
of the anomalous acceleration of about 1 nm s$^{-2}$ experienced
by the Pioneer 10/11 spacecraft in the range $20$ AU$\lesssim
r\lesssim 70$ AU (Anderson et al. 1998; 2002).

Basically, Jaekel and Reynaud (2006) start from a space-time line
element \eqi
ds^2=g_{00}(r)c^2dt^2+g_{rr}(r)(dr^2+r^2d\theta^2+r^2\sin^2\theta
d\phi^2 ),\eqf written in the standard Eddington isotropic
coordinates, and write the metric coefficients as sums of standard
relativistic expressions and small deviations \eqi
g_{\mu\nu}\equiv[g_{\mu\nu}]_{\rm st}+\delta g_{\mu\nu}, \ |\delta
g_{\mu\nu} |\ll 1.\eqf The two sectors $\delta g_{00}(r)$ and
$\delta(g_{00}g_{rr})(r)$ yield two anomalous potentials
$\delta\Phi_{N}$ and $\delta\Phi_{P}$ which affect the motion of
test particles and light rays.

In this note we do not demand to discuss the model proposed by
Jaekel and Reynaud (2006) in all of its generality, but only as
far as  possible explanations of the Pioneer anomaly are
concerned.

\section{On the first anomalous potential}
In regard to the correction $\delta\Phi_N$ to the
Newtonian potential
 coming from the first sector, Jaekel and Reynaud
(2006) write: ``Should the Pioneer anomaly be explained by an
anomaly in the first sector, a linear dependence of the potential
$\delta\Phi_{ N}$ would be needed to reproduce the fact that the
anomaly has a roughly constant value over a large range of
heliocentric distances $r_{\rm P}$ \eqi
c^2\partial_r\delta\Phi_N\simeq a_P,\ 20\ {\rm AU}\leq r_P\leq 70\
{\rm AU}. \lb{jrcaz1}\eqf The simplest way to modelize the anomaly
would thus correspond to a potential varying linearly with $r$ and
vanishing at Earth orbit [...]" In regard to the compatibility of
such an extra-potential with the observed features of the
planetary motions, especially in the regions in which the Pioneer
anomaly manifested itself, according to our present-day knowledge
of it, the predicted action of an anomalous constant and uniform,
radial acceleration of about 1 nm s$^{-2}$ on the orbits of the
outer planets of the Solar System was investigated in the
framework of  the latest observations in a number of papers always
getting neat and unambiguous negative answers.

Iorio and Giudice (2006) compared the time-dependent patterns of
the directly observable quantities $\alpha\cos\delta$ and
$\delta$, where $\alpha$ and $\delta$ are the planetary right
ascension and the declination, respectively, induced by a
Pioneer-like extra-acceleration for Uranus (19.19 AU), Neptune
(30.06 AU) and Pluto (39.48 AU) to their observational residuals
obtained by processing almost one century (1913-2003) of optical
data with the RAS IAA EPM2004 ephemerides (Pitjeva 2005). Tangen
(2006) did the same by using a different theoretical quantity.
While a Pioneer-type force would affect $\alpha\cos\delta$ and
$\delta$ with polynomial signatures of hundreds of arcseconds, the
observed residuals are almost uniform strips well constrained
within $\pm 5$ arcseconds. It is interesting to note that the very
same conclusion could already have been traced long time ago  by
using the residuals of some sets of modern optical observations
(1984-1997) to the outer planets processed by Morrison and Evans
(1998) with the  NASA JPL DE405 ephemerides. Analysis of residuals
obtained with even older ephemerides would have yielded the same
results. Foe example, Standish (1993) used JPL DE200 ephemerides
to process optical data of Uranus dating back to 1800: the
obtained residuals of $\alpha$ and $\delta$ do not show any
particular structure being well constrained within $\pm 5$
arcseconds. Gomes and Ferraz-Mello (1987) used the VSOP82
ephemerides to process more than one century (1846-1982) of
optical data of Neptune getting no anomalous signatures as large
as predicted by the presence of a Pioneer-like anomalous force. In
regard to Pluto, Gemmo and Barbieri (1994) and Rylkov et al.
(1995) used the JPL DE200 and JPL DE202 ephemerides in  producing
residuals of $\alpha$ and $\delta$: no Pioneer-type signatures can
be detected in them.

Pitjeva (2006) recently determined the anomalous secular rates of
perihelion $\varpi$ for Jupiter (5.2 AU), Saturn (9.5 AU) and
Uranus by contrasting, in a least-square sense\footnote{Contrary
to the right ascension and declination, the perihelia are not
directly observable.}, almost one century of mainly optical (apart
from Jupiter) data with the full model of relevant Newtonian and
Einsteinian dynamical effects of the not yet released EPM2006
ephemerides. After comparing them with the theoretical predictions
for such precessions induced by a Pioneer-like acceleration, we
got another clear negative answer, as pointed out in  (Iorio
2006a; 2006b).

Even the use of the Voyager 2 radio-tracking data to Neptune ruled
out the existence of a Pioneer-like acceleration  which would
affect the Neptune semi-major axis with a totally undetected
short-period effect (Iorio 2006c; 2006b).

It is important to stress that such conclusions are purely
phenomenological and model-independent: no speculations about the
possible origin of such an extra-acceleration at all have been
used.

In conclusion, we cannot agree with Jaekel and Reynaud (2006) when
they write: ``[...] it then remains to decide whether or not the
ephemeris of the outer planets are accurate enough to forbid the
presence of the linear dependence (\ref{jrcaz1}) in the range of
distances explored by the Pioneer probes (Iorio and Giudice 2006;
Tangen 2006). This point remains to be settled (Brownstein and
Moffat 2006)''. It is just the case to note that, in fact, the
gravitational mechanism put forth by Brownstein and Moffat (2006)
by fitting all the presently available Pioneer 10/11 data to a
Yukawa-type model\footnote{Instead, Jaekel and Reynaud (2006)
write that ``Brownstein and Moffat have explored the possibility
that the linear dependence holds at distance explored by Pioneer
probes [...]" } completely failed when applied to the perihelia of
Jupiter, Saturn and Uranus (Iorio 2006d; 2006b).

\section{On the second anomalous potential}
\begin{figure}
\begin{center}
\includegraphics[width=14cm,height=11cm]{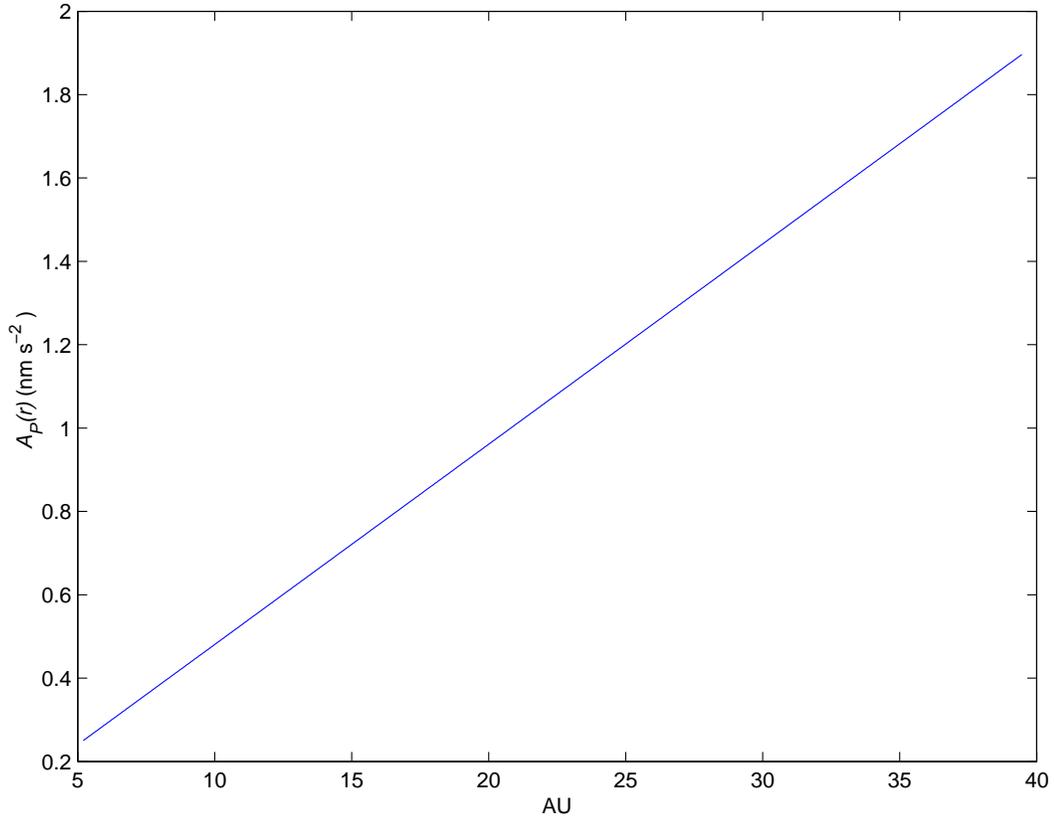}
\end{center}
\caption{\label{JRquad} Anomalous acceleration, in nm s$^{-2}$,
induced by $\delta\Phi_P=c^2\chi r^2$, with $\chi=4\times 10^{-8}$
AU$^{-2}$, according to Jaekel and Reynaud (2006). }
\end{figure}

In regard to the second sector, Jaekel and Reynaud (2006) write:
``In any case, there is another possibility, namely that the
Pioneer anomaly is induced by the second anomalous potential
$\delta\Phi_P$ rather than the first one $\delta\Phi_N$. We now
consider these terms which are still here even if there is no
anomaly at all in the first sector ($\delta\Phi_N=0$)." As a
result of their investigation, Jaekel and Reynaud (2006) find
that: ``A roughly constant anomaly is produced when [...]
$\delta\Phi_P(r)$ is quadratic in $r$, in the range of Pioneer
distances." Their choice is \eqi\delta\Phi_P(r)=c^2\chi r^2,\
\chi\simeq 4\times 10^{-8}\ {\rm AU}^{-2}, \eqf where  $c$ is the
speed of light in vacuum. The resulting acceleration \eqi
A_P(r)=-2c^2\chi r\lb{accnew},\eqf in units of nm s$^{-2}$, is
plotted in Figure \ref{JRquad}.
Without investigating how well such a model fits, in fact, all 
the currently available data of the Pioneer 10/11 spacecraft, here
we are going to derive theoretical predictions for the secular
perihelion advance induced by \rfr{accnew}. The standard methods
of perturbative celestial mechanics yield \eqi\dert\varpi
t=-3c^2\chi\sqrt{ \rp{a^3(1-e^2)}{GM} },\lb{perinew}\eqf where $a$
and $e$ are the semi-major axis and the eccentricity,
respectively, of the planet's orbit, $G$ is the Newtonian
gravitational constant and $M$ is the mass of the Sun. Note that
\rfr{perinew} is an exact result. The comparison among the
anomalous advances for Jupiter, Saturn and Uranus predicted with
\rfr{perinew} and the determined perihelia rates is in Table
\ref{tavola}.
{\small\begin{table}\caption{ First row: determined
extra-precessions of the perihelia of Jupiter, Saturn and Uranus,
in arcseconds per century (Pitjeva 2006). The quoted uncertainties
are the formal, statistical errors re-scaled by a factor 10 in
order to get realistic estimates. Second row: predicted anomalous
extra-precessions of the perihelia for Jupiter, Saturn and Uranus,
in arcseconds per century, according to \rfr{perinew}.
}\label{tavola}

\begin{tabular}{llll} \noalign{\hrule height 1.5pt}

 & Jupiter & Saturn  & Uranus\\
$\dot\varpi_{\rm meas}$ & $0.0062\pm 0.036$ & $-0.92\pm 2.9$ & $0.57\pm 13.0$\\
$\dot\varpi_{\rm pred}$ & -18.679 & -46.3 & -132.3\\
\hline

\noalign{\hrule height 1.5pt}
\end{tabular}

\end{table}}

As can be noted, even by re-scaling by a factor 10 the formal
errors released by Pitjeva (2006), the discrepancy among the
predicted and the determined values amounts to 519, 15 and 10
sigma for Jupiter, Saturn and Uranus, respectively.

\section{Conclusions}
In this note we investigated the new proposal by Jaekel and
Reynaud (2006) to accommodate the Pioneer anomaly in the framework
of their post-Einsteinian metric extension of general relativity.
First, we reviewed the wealth of observational evidence pointing
against the presence in planetary data of any anomalous signature
as large as predicted by an anomalous Pioneer-type acceleration
which could, e.g., be induced by an extra-potential linear with
distance like $\delta\Phi_N$ by Jaekel and Reynaud (2006). We not
only used the RAS IAA EPM2004 ephemerides, as already done in
previous works, but the NASA JPL DE405, DE200 and DE202
ephemerides and the VSOP82 theory as well. This should be
sufficient to rule out, among other things, the presence of the
first anomalous potential $\delta\Phi_N$ of the Jaekel and Reynaud
(2006) model. Their second anomalous potential $\delta \Phi_P$,
which would be able to reproduce the behavior of the Pioneer
probes by assuming a quadratic dependence with distance, would
also affect the orbital motion of the planets of the Solar System
with extra-perihelion rates of tens-hundreds of arcseconds per
century, at least in the regions in which the Pioneer anomaly
manifested itself in its presently known form. Anomalous
perihelion precessions of so large size are completely ruled out
by the latest RAS IAA determinations of the perihelion rates for
Jupiter, Saturn and Uranus by more than 10 sigmas, even after
re-scaling by a factor 10 the formal errors released by Pitjeva.




\end{document}